\documentclass[twocolumn]{aastex6}

\usepackage{amssymb}
\usepackage{amsfonts}
\usepackage{amsmath}
\usepackage[varg]{txfonts}
\usepackage{natbib}
\usepackage{color}
\usepackage{graphicx}
\usepackage{hyperref}

\newcommand{\vect}[1]{\boldsymbol{#1}}

\shorttitle{Neutron star kicks in asymmetric supernova explosions}

\shortauthors{H.-Thomas Janka}

\begin{document}

\title{Neutron star kicks by the gravitational tug-boat 
mechanism in asymmetric supernova explosions:
progenitor and explosion dependence}

\author{Hans-Thomas Janka\altaffilmark{1}}
\altaffiltext{1}{Max-Planck-Institut f\"ur Astrophysik,
       Karl-Schwarzschild-Str.~1, D-85748 Garching, Germany}

\begin{abstract}
Asymmetric mass ejection in the early phase of supernova (SN)
explosions can impart a kick velocity to the new-born neutron
star (NS). For neutrino-driven explosions the NS acceleration 
was shown to be mainly caused by the gravitational attraction 
of the anisotropically expelled inner ejecta, while hydrodynamic 
forces contribute on a subdominant level, and asymmetric neutrino 
emission plays only a secondary role. Two- and three-dimensional
hydrodynamic simulations demonstrated that this gravitational
tug-boat mechanism can explain the observed space velocities 
of young NSs up to more than 1000\,km\,s$^{-1}$. Here, we 
discuss how the NS kick depends on the energy, ejecta
mass, and asymmetry of the SN explosion, and which role the
compactness of the pre-collapse stellar core plays for the 
momentum transfer to the NS. We also provide simple 
analytic expressions for the NS velocity in terms of these
quantities. Referring to results of hydrodynamic simulations
in the literature, we argue why within the discussed scenario
of NS acceleration, electron-capture SNe, low-mass Fe-core SNe and
ultra-stripped SNe can be expected to have considerably lower
intrinsic NS kicks than core-collapse SNe of massive stellar cores.
Our basic arguments remain valid also if progenitor stars
possess large-scale asymmetries in their convective 
silicon and oxygen burning layers. Possible scenarios
for spin-kick alignment are sketched. Much of our discussion
stays on a conceptual and qualitative level, and more
work is necessary on the numerical modeling side to determine
the dependences of involved parameters, whose prescriptions
will be needed for recipes that can be used to better describe
NS kicks in binary evolution and population synthesis studies.
\end{abstract}


\section{Introduction}

Neutron stars (NSs) are born with kick velocities of typically
200--500\,km\,s$^{-1}$, which is evidenced by the measured 
proper motions of young radio pulsars (exceeding the break-up
velocities of close double-star systems) and by the orbital 
parameters and spin orientations of NSs in binary systems 
\citep[e.g.,][]{Harrisonetal1993,Kaspietal1996,LyneLorimer1994,Fryeretal1998,Laietal2001,Arzoumanianetal2002,Chatterjeeetal2005,Hobbsetal2005}.
Also recently detected hypervelocity stars might originate
from disrupted binaries via asymmetric supernovae (SNe) with
large NS kicks \citep{Tauris2015}.
These NS kicks could be a consequence of asymmetric 
explosions \citep[e.g.,][]{JankaMueller1994,BurrowsHayes1996}
or anisotropic emission of the neutrinos that carry
away the huge binding energy of the compact star 
\citep[e.g.,][]{Woosley1987,Bisnovatyi-Kogan1993,FryerKusenko2006,Kusenkoetal2008,SagertSchaffner2008}.

The non-radial
flow instabilities \citep[convective overturn and the standing
accretion shock instability, SASI;][]{Blondinetal2003,Foglizzo2002,Foglizzoetal2006,Foglizzoetal2007,Schecketal2008} 
that develop shortly after core bounce in the postshock accretion
layer of collapsing stellar cores, produce mass-ejection asymmetries
of the SN explosion that can be sufficiently large to account for
NS kicks of several hundred km\,s$^{-1}$ with cases reaching
up to and even beyond 1000\,km\,s$^{-1}$. This was demonstrated by 
two-dimensional (2D) and three-dimensional (3D)
hydrodynamic simulations of neutrino-driven SN explosions
\citep{Schecketal2004,Schecketal2006,Nordhausetal2010,Nordhausetal2012,
Wongwathanaratetal2010,Wongwathanaratetal2013}, which led to
the conclusion, consistent with linear momentum conservation, that
the NS must receive a natal kick opposite to the total momentum of the 
SN ejecta.

Recently \citet{BrayEldridge2016} applied such a
connection of NS kicks and SN explosion asymmetries, following
their inspiration and motivated by the interpretation
of observational and modeling results of Cassiopeia~A.
Correspondingly, they suggested a direct relationship between the 
velocity of the compact remnant and the ratio of SN ejecta mass to
NS mass \citep[this assumption was also made by][]{Beniaminietal2016}.
It is important to note that such a relationship would not necessarily
hold if NS kicks were caused by anisotropic neutrino emission
\citep[e.g.][]{FryerKusenko2006}.

During the asymmetric ejection of the SN debris, the NS acceleration
happens by momentum transfer through hydrodynamic pressure forces
and momentum advection in outflows and downflows, by gravitational
forces of the anisotropic ejecta on the compact remnant, and by
nonspherical neutrino-emission. \citet{Schecketal2006} and
\citet{Wongwathanaratetal2013} found that the anisotropic 
gravitational interaction as a non-saturating long-range force
has significant influence over several seconds and therefore
is by far the most dominant effect, whereas neutrino emission
associated with asymmetric accretion contributes only on a low 
level. For this reason \citet{Wongwathanaratetal2013}
introduced the term ``gravitational tug-boat mechanism''
for the physical process that is mainly responsible for the
NS recoil associated with the asymmetric ejection of matter in
SN explosions: the slowest and usually densest and most massive
ejecta ``clumps'' exert the strongest forces on the NS such
that their pull accelerates the compact remnant opposite to 
the direction of the more powerful explosion. This 
scenario is consistent with the conservation of the
total linear momentum of SN ejecta and NS in the rest-frame
of the progenitor star. Since the NS is accelerated in the
direction opposite to the stronger explosion and over
a time scale of several seconds (which is longer than the
explosive iron-group nucleosynthesis), large NS kicks are 
predicted to correlate with considerably more production of 
elements from silicon to the iron group in the hemisphere
pointing away from the NS kick vector \citep{Wongwathanaratetal2013}.
This prediction seems to be fully compatible with the NuSTAR
map of the $^{44}$Ti distribution in Cassiopeia~A 
\citep{Wongwathanaratetal2016}.

If high birth kicks of black holes
(BHs) are required to explain the spatial distribution of 
Galactic low-mass X-ray binaries containing BHs
(\citealt{Repettoetal2012,RepettoNelemans2015}; for 
counter-arguments, however, see \citealt{Mandel2016}), the 
gravitational tug-boat mechanism can offer a scenario in
which the high kicks are a consequence of considerable 
amounts of matter that remain gravitationally bound during
the SN explosion and fall back asymmetrically to the compact
remnant \citep{Janka2013}. Otherwise, if high natal BH kicks 
are not required by observations, the operation of the 
gravitational tug-boat mechanism would mean that BH formation
events with significant SN fallback are rare in the Galactic
neighborhood, compatible with conclusions drawn from recent 
theoretical studies of the progenitor-SN connection
\citep{Uglianoetal2012,Ertletal2016,Sukhboldetal2016}. Or,
alternatively, it could mean that most BHs originate from 
SNe with very little or rather spherical mass ejection.

The majority of cases simulated by 
\citet{Schecketal2006} and \citet{Wongwathanaratetal2013} 
produced NS kicks of several 100\,km\,s$^{-1}$ in good match
with the maximum of the observed NS velocity distribution.
In the set of 
2D models of \citet{Schecketal2006}, one out of 70 cases developed
an estimated NS kick velocity of more than 1000\,km\,s$^{-1}$. In 3D
such extreme velocities have not been obtained yet, but the simulations
by \citet{Wongwathanaratetal2013} were constrained to a small sample
of progenitor stars and a set of only 20 models. One of these cases
showed a NS kick velocity in excess of 700\,km\,s$^{-1}$ after
3.3\,s of post-bounce evolution, with an acceleration of more
than 70\,km\,s$^{-2}$ still boosting the NS velocity. Three other
cases had NS kick velocities of nearly 600\,km\,s$^{-1}$ at 3.3\,s
after bounce with ongoing accelerations of up to $>$100\,km\,s$^{-2}$.
Velocities
above 1000\,km\,s$^{-1}$, maybe even considerably exceeding this
value, seem to be well possible when the incipient explosion develops a large
dipolar asymmetry mode, or when a long-lasting phase of asymmetric accretion
transfers momentum to the nascent NS, both of which cases did not
occur in the limited set of 3D models 
computed by \citet{Wongwathanaratetal2013}.
Although such scenarios are plausible, given the large explosion
asymmetries seen in recent 3D SN models
\citep[e.g.,][]{Melsonetal2015,Lentzetal2015,Mueller2016}, a direct
numerical demonstration of NS kicks beyond 1000\,km\,s$^{-1}$ by
the gravitational tug-boat mechanism in 3D SN simulations is desirable.

The papers by \citet{Schecketal2006} and \citet{Wongwathanaratetal2013}
investigated different progenitors with explosion
energies and asymmetries varying over rather wide ranges,
but the small variety of 15\,$M_\odot$ and 20\,$M_\odot$ models
did not allow them to 
illuminate the systematics of NS kicks in dependence on
stellar progenitor properties and SN explosion properties.
In this work we attempt to
take first steps in this direction. Our discussion will
remain mostly on a conceptual and qualitative level,
focussing on scaling-laws derived here to provide insights into
basic factors that determine the NS kicks by the gravitational
tug-boat mechanism. We thus intend to prepare the ground for
future comparisons of the theoretical predictions
with observations and for an improved description
of NS kicks in binary evolution and population synthesis studies.
For these future goals to become achievable, the parameters appearing
in our scaling laws need to be pinned down in their dependences
by more elaborate and longer 3D simulations of larger sets of
progenitor stars than currently available in the literature.
In the present paper we will briefly review the published results 
and will interpret their meaning and shortcomings.

The paper is structured as follows. In Sect.~\ref{sec:expdept}
we present simple scaling relations for the NS kick
velocity as a function of the mass of neutrino-heated ejecta,
explosion energy, and explosion asymmetry, 
in Sect.~\ref{sec:progdept} we will analyse the role of the
core structure and compactness of the progenitor stars, in 
Sect.~\ref{sec:numerical} we will connect the parameters in
the scaling laws to simulation results in the literature,
and in Sect.~\ref{sec:spinkick} we will discuss possibilities
to explain a putative spin-kick alignment suggested by observations.
Sect.~\ref{sec:conclusions} contains a summary and conclusions.

\section{Analytical scaling relations}
\label{sec:results}

How does the kick velocity of new-born NSs depend on the 
structure
of the progenitor stars and on the characteristic properties 
of the SN explosions? This question is not only relevant for
interpreting observations, e.g.\ of hypervelocity stars 
\citep{Tauris2015}, but also, for example,
for stellar population studies and for understanding the 
evolution of binary stars that give birth to double NS systems
\citep[e.g.][]{VossTauris2003,Taurisetal2016}.
In theoretical models for population synthesis a widely used
approach to implement the effects of NS natal kicks is based on
single-component or multi-component Maxwell-Boltzmann velocity 
distributions 
\citep[see, e.g.,][]{VossTauris2003,Dominiketal2015},
which was questioned recently by \citet{BrayEldridge2016}.
Instead, the latter authors proposed a simple relationship for
the NS kick velocity defined as a linear function of the ratio
of the SN ejecta mass to the remnant mass.

Though this suggestion seems to work well, no detailed
physical explanation was provided in \citet{BrayEldridge2016}.
In the following we will provide corresponding arguments and
will, based on our current understanding of neutrino-driven SN
explosions and the gravitational tug-boat mechanism for 
NS acceleration, derive simple scaling relations for the
NS kick velocity, $v_\mathrm{NS}$, as function of basic parameters
that are linked to the ejecta mass of the SN, the mass-ejection
asymmetry, and the energy of the explosion. We will also discuss
how the efficieny of the NS acceleration mechanism depends on
the density profile and compactness of the stellar core 
above the initial mass cut, from where the SN shock starts
its outward expansion. 
Our discussion will stay mostly on a didactic and qualitative
level, because, as we will argue in Sect.~\ref{sec:numerical},
hydrodynamic SN explosion simulations are needed to determine
the values of the parameters that occur in our formulas.
The simulation results presently available in the literature,
however, only provide crude guidance and do not allow for
definitive, finally quantitative conclusions on the general
dependences on progenitor and explosion properties. 
The only exception is a distinct difference between the NS
kicks that can be expected for stars near the low-mass end of
the SN progenitors and some lighter cases of ultra-stripped SNe
on one side, and SNe of progenitors with massive iron cores
at collapse on the other side.

\subsection{Kick dependence on the explosion properties}
\label{sec:expdept}

Considering only the NS kick associated with anisotropic
mass ejection, momentum conservation in the frame of the 
progenitor star implies that NS and ejecta momenta are
equal in value and opposite in direction, i.e.:
\begin{equation}
v_\mathrm{NS} = |\vect{v}_\mathrm{NS}| = 
\alpha_\mathrm{ej}\, P_\mathrm{ej}\, M_\mathrm{NS}^{-1} \,,
\label{eq:vns}
\end{equation}
where $M_\mathrm{NS}$ is the mass of the NS.\footnote{While
one is tempted to interpret $M_\mathrm{NS}$ in this equation
as the gravitational mass of the NS, the velocity estimate 
is better in fact if the NS mass is taken to be the baryonic
mass, because much of the hydrodynamic recoil momentum is
imparted to the NS at a time when the compact remnant has
not yet lost a major fraction of its gravitational 
binding energy by neutrinos. 
Note, however, that all of our estimates presented in
this paper are not on a level of accuracy that requires a
very careful distinction between baryonic and
gravitational NS mass, which makes a difference of order
$\sim$10\% only.}
$P_\mathrm{ej}$ is defined by the volume integral
\begin{equation}
P_\mathrm{ej} = \int_{M_\mathrm{ej}}\mathrm{d}V\,\rho\,|\vect{v}|\,,
\label{eq:pej}
\end{equation}
with $M_\mathrm{ej}$ being the relevant ejecta mass
to be further discussed below. In Eq.~(\ref{eq:vns})
we introduce the momentum-asymmetry parameter
\begin{equation}
\alpha_\mathrm{ej} = \frac{|\vect{P}_\mathrm{gas}|}{P_\mathrm{ej}}\,,
\label{eq:alpha}
\end{equation}
when the momentum integral of the ejecta gas is calculated as
\begin{equation}
\vect{P}_\mathrm{gas} = 
\int_{M_\mathrm{ej}}\mathrm{d}V\,\rho\,\vect{v}\,.
\label{eq:pgas}
\end{equation}
By its definition, $P_\mathrm{ej}$ is related to the kinetic
energy of the ejecta: $E_\mathrm{kin} = \frac{1}{2}M_\mathrm{ej}
\bar{v}_\mathrm{ej}^2 = \frac{1}{2}P_\mathrm{ej}^2 M_\mathrm{ej}^{-1}$,
where $\bar{v}_\mathrm{ej}$ is the average ejecta velocity.
This yields
\begin{equation}
P_\mathrm{ej} = \sqrt{2\,E_\mathrm{kin}\,M_\mathrm{ej}}
=  \sqrt{2\,f_\mathrm{kin}E_\mathrm{exp}\,M_\mathrm{ej}} \,,
\label{eq:pej2}
\end{equation}
where in the second expression we introduced the parameter 
$f_\mathrm{kin}$ to relate the kinetic energy of the explosion
during the relevant time of NS acceleration with the final 
SN explosion energy $E_\mathrm{exp}$. 

Both the values of $\alpha_\mathrm{ej}$ and $f_\mathrm{kin}$
are time-dependent and need to be determined by multi-dimensional
hydrodynamic simulations.

Using Eq.~(\ref{eq:pej2}) in Eq.~(\ref{eq:vns}), we can derive
the expression
\begin{eqnarray}
v_\mathrm{NS} &=& \alpha_\mathrm{ej}\,
\sqrt{2\,f_\mathrm{kin}E_\mathrm{exp}\,M_\mathrm{ej}}\,M_\mathrm{NS}^{-1}
\label{eq:vns2a}\\
&=& 211\,\mathrm{km\,s}^{-1}\,f_\mathrm{kin}^{1/2}
\,\frac{\alpha_\mathrm{ej}}{0.1}\,
\left(\frac{E_\mathrm{exp}}{10^{51}\,\mathrm{erg}}\right)^{1/2} 
\times\nonumber \\
&\phantom{=}& 
\phantom{211\,\mathrm{km\,s}^{-1}\,f_\mathrm{kin}^{1/2}
\,\frac{\alpha_\mathrm{ej}}{0.1}\,}
\!\!\!\!\!\!\!\times\,\,
\! \left(\frac{M_\mathrm{ej}}{0.1\,M_\odot}\right)^{1/2}
\! \left(\frac{M_\mathrm{NS}}{1.5\,M_\odot}\right)^{-1} .
\label{eq:vns2b}
\end{eqnarray}

After the NS kick has saturated \citep[typically after a few
seconds, see][]{Schecketal2006,Wongwathanaratetal2013}, 
the NS momentum, 
$P_\mathrm{NS} = M_\mathrm{NS}v_\mathrm{NS}$, and the ejecta
momentum, $|\vect{P}_\mathrm{gas}|=\alpha_\mathrm{ej}P_\mathrm{ej}$,
remain constant, but $\alpha_\mathrm{ej}$, $E_\mathrm{kin}$,
and $M_\mathrm{ej}$ may still evolve. We therefore consider as
the relevant mass of $M_\mathrm{ej}$ (determining the time
when $\alpha_\mathrm{ej}$ is measured and vice versa) 
the mass that is accumulated behind the outgoing SN shock until
the NS kick asymptotes to its final value. The representative
values chosen for normalizing the quantities in Eq.~(\ref{eq:vns2b}) 
are guided by the results of \citet{Schecketal2006} and
\citet{Wongwathanaratetal2013} (see tables and plots there), 
where they are typical of
exploding models after about one second of post-bounce
evolution (i.e., near the end of the phase of explosive
nucleosynthesis in shock-heated ejecta). As the shock propagates
farther outward and sweeps up more matter from the spherical
progenitor star, the ejecta mass increases while the value of
the asymmetry parameter decreases, but the product of both
quantities remains constant (unless anisotropic fallback
modifies the ejecta asymmetry). It is important to note that,
if $M_\mathrm{ej}$ in Eqs.~(\ref{eq:vns2a}) and (\ref{eq:vns2b})
were to be interpreted as the total ejecta mass of the SN, 
$\alpha_\mathrm{ej}$ would have to be measured
after the shock-breakout from the surface of the
progenitor star, i.e., after the blast wave has accelerated
all of the outer stellar layers.
This is not very practical in numerical simulations of the 
explosion mechanism and NS acceleration, which are usually
only carried over a few seconds at most.

In the case of neutrino-driven SN explosions, the ejecta mass
that is relevant for the NS acceleration is tightly correlated
with the amount of matter that is advected through the shock and 
accreted towards the nascent NS in downflows
to be neutrino heated near the gain radius and anisotropically 
expelled again. Energy absorption from neutrinos lifts this
matter to a state of neutral gravitational binding, and the
recombination energy released when free nucleons assemble to
$\alpha$-particles and heavy nuclei in the outflow provides a
positive contribution of typically $\epsilon \sim (5\,...\,8)\,
\mathrm{MeV/nucleon}\approx (5\,...\,8)\times 10^{18}$\,erg\,g$^{-1}$
to the explosion energy of the SN 
\citep{Janka2001,Schecketal2006,MarekJanka2009,Mueller2015}.
Ignoring the additional positive energy contributions from
explosive nuclear burning in shock-heated ejecta and from the
(essentially spherical) neutrino-driven wind that follows 
the transient post-bounce phase of simultaneous accretion and 
asymmetric mass (re-)ejection, and also ignoring the negative 
energy of the gravitational binding of the overlying stellar
layers ahead of the SN shock, we can (approximately) write:
\begin{equation}
E_\mathrm{exp} \approx \epsilon\,M_{\mathrm{ej},\nu} = 
\epsilon\,\beta_\nu\,M_\mathrm{ej}  \,.
\label{eq:experg}
\end{equation}
where $M_{\mathrm{ej},\nu}$ defines the mass of neutrino-heated
postshock ejecta, which we relate to the total (expelled) 
postshock mass by the parameter $\beta_\nu\le 1$:
$M_{\mathrm{ej},\nu}=\beta_\nu M_\mathrm{ej}$.
Equation~(\ref{eq:experg}) expresses the fact that in the 
neutrino-driven mechanism the mass of the neutrino-heated ejecta
determines the energy of the explosion.
Considering $\epsilon$ roughly as a constant, 
Eq.~(\ref{eq:experg}) implies a linear relation between
$E_\mathrm{exp}$ and $M_{\mathrm{ej},\nu}$,
which is supported by large sets of 2D explosion simulations
(\citealt{Schecketal2006}, see figures~9 and 10 and 
Appendix~C there; 
\citealt{Gessner2014,GessnerJanka2017}).\footnote{Figure~9
in \citet{Schecketal2006} does not directly display the linear
dependence between $E_\mathrm{exp}$ and $M_{\mathrm{ej},\nu}$,
but instead it shows a linear increase of $E_\mathrm{exp}$ and
a linear decrease of $M_\mathrm{NS}$ with the boundary neutrino
luminosity $L_\mathrm{ib}$, which is a governing parameter that
regulates the explosion energy of the explosion simulations. 
At this point it is important to recall the fact that 
the mass of the neutrino-heated ejecta, $M_{\mathrm{ej},\nu}$,
is given by the mass in the gain layer at the time of the onset
of the explosion ($\Delta M_\mathrm{gain}(t_\mathrm{exp})$ in
figure~C.1 of \citealt{Schecketal2006}) plus the additional
mass that the neutrino-driven wind supplements to the
neutrino-heated ejecta ($M_\mathrm{wind}$) after the onset of 
the explosion: $M_{\mathrm{ej},\nu} = 
\Delta M_\mathrm{gain}(t_\mathrm{exp}) + M_\mathrm{wind}$.
Because the neutrino-driven wind feeds this mass
into the ejecta at the expense of the NS mass, the wind mass 
increases proportionally to the decrease of $M_\mathrm{NS}$,
i.e.\ $M_\mathrm{wind} \propto M_\mathrm{NS}^0-M_\mathrm{NS}
\propto L_\mathrm{ib} \propto E_\mathrm{exp}$ (the proportionalities
can be concluded from figure~9 of \citealt{Schecketal2006}; 
$M_\mathrm{NS}^0$ is an upper bound on the baryonic NS mass,
whose exact value is not relevant for the argument).
Also $\Delta M_\mathrm{gain}(t_\mathrm{exp})$ increases
linearly with the explosion energy,
$\Delta M_\mathrm{gain}(t_\mathrm{exp})\propto E_\mathrm{exp}$,
which can be seen by combining the linear relations of 
figures~C.1 and C.2 or those of figures~10 and 9
of \citet{Schecketal2006}. Taking all this information
together one verifies the desired linear dependence
$M_{\mathrm{ej},\nu} =
\Delta M_\mathrm{gain}(t_\mathrm{exp}) + M_\mathrm{wind}
\propto E_\mathrm{exp}$. Similar arguments apply for the
interpretation of the results shown in \citet{Gessner2014}.}
Using Eq.~(\ref{eq:experg}) with this assumption
in Eq.~(\ref{eq:pej2}), and assuming also 
$\beta_\nu\sim\mathrm{const}$, we obtain the proportionality 
relations
\begin{equation}
P_\mathrm{ej} \propto M_\mathrm{ej} \propto E_\mathrm{exp}\,,
\label{eq:pej3}
\end{equation}
which are nicely confirmed by results of multi-dimensional
SN simulations during the first second(s) of the explosion
(\citealt{Schecketal2006}, figure~11; \citealt{Gessner2014}).

Employing a typical value of $\epsilon\sim 5\,$MeV/nucleon
in Eq.~(\ref{eq:experg}), which implies
\begin{equation}
\frac{\beta_\nu\,M_\mathrm{ej}}{0.1\,M_\odot}\approx\epsilon_5^{-1}\,
\frac{E_\mathrm{exp}}{10^{51}\,\mathrm{erg}} \,,
\label{eq:emrelation}
\end{equation}
where $\epsilon_5 = \epsilon/(5\,\mathrm{MeV/nucleon})$, 
we can replace $M_\mathrm{ej}$ in Eq.~(\ref{eq:vns2b}) by 
$E_\mathrm{exp}$ to get
\begin{eqnarray}
v_\mathrm{NS} 
&=& 211\,\mathrm{km\,s}^{-1}\,
\left(\frac{f_\mathrm{kin}}{\epsilon_5\,\beta_\nu}\right)^{\! 1/2}
\left(\frac{\alpha_\mathrm{ej}}{0.1}\right)\,\times \nonumber\\
&\phantom{=}&
\phantom{211\,\mathrm{km\,s}^{-1}\,
\left(\frac{f_\mathrm{kin}}{\epsilon_5\,\beta_\nu}\right)^{\! 1/2}}
\!\!\!\!\!\!\!\times\,
\left(\frac{E_\mathrm{exp}}{10^{51}\,\mathrm{erg}}\right)
\left(\frac{M_\mathrm{NS}}{1.5\,M_\odot}\right)^{-1} .
\label{eq:vns3}
\end{eqnarray}

Equation~(\ref{eq:vns3}) is the main result of this paper.
It means that the NS kick grows roughly linearly with the 
explosion energy (or, alternatively, with the relevant
ejecta mass, $M_\mathrm{ej}$, by means of 
Eq.~\ref{eq:emrelation}) and with the explosion asymmetry 
$\alpha_\mathrm{ej}$. Both dependences are easy to 
understand: a more asymmetric and more powerful explosion
is able to impart a larger kick to the NS.
The parameters $f_\mathrm{kin}$ and $\beta_\nu$ depend on 
the SN shock dynamics and therefore on the radial structure 
of the progenitor star and the evolution stage of the SN
explosion. However, the combination of parameters in the 
factor $f_\mathrm{kin}/(\epsilon_5\,\beta_\nu)$ is typically
of order unity at the time when the NS kick saturates
(since $f_\mathrm{kin}$ and $\beta_\nu$ are
of similar magnitude), and its variation is moderated by
the square root of this factor in Eq.~(\ref{eq:vns3}).
The momentum asymmetry parameter $\alpha_\mathrm{ej}$ depends
on the stochastic growth of hydrodynamic instabilities in the
postshock layer, which trigger the onset of an asymmetric
explosion. Kick velocities in excess of 1000\,km\,s$^{-1}$
require $\alpha_\mathrm{ej} \gtrsim 0.5$ for all other 
factors in Eq.~(\ref{eq:vns3}) being unity, which is within
reach of some published explosion models 
\citep[e.g.,][]{Schecketal2006,Wongwathanaratetal2013}.
Of course, $f_\mathrm{kin}$, $\beta_\nu$, $\epsilon$,
and $\alpha_\mathrm{ej}$ could contain hidden dependences on
$E_\mathrm{exp}$ and the progenitor star, which can only be 
determined by hydrodynamic explosion modeling.

For practical applications of Eqs.~(\ref{eq:vns2b}) or 
(\ref{eq:vns3}), for example in population synthesis studies,
it may be preferable to express the NS kick velocity in
terms of the total ejecta mass of the SN, $M_\mathrm{ej,SN}$ 
\citep[see][]{BrayEldridge2016}, or the NS mass. In this context
it is worth noting that on grounds of their semi-analytic 
model for the SN progenitor-explosion connection,
\citet{Muelleretal2016b} suggested loose correlations between the
SN explosion energy, the total ejecta mass of the SN, and the
(gravitational) NS mass, which they interpreted as compatible
with power-law relations deduced from observational analyses
(see their figures~9 and 11). The scatter of their model
data, however, is large, and the correlations of the 
mentioned quantities are much less clearly defined than the 
tight relation between $v_\mathrm{NS}$ and $E_\mathrm{exp}$
expressed by Eq.~(\ref{eq:vns3}). Similarly, there is an 
approximately linear relation between the SN explosion energy 
and the ejected $^{56}$Ni mass (see figure~17 in 
\citealt{Sukhboldetal2016} and, for the iron-group mass as a
proxy of the $^{56}$Ni mass, figure~10 in 
\citealt{Muelleretal2016b}). By means of Eq.~(\ref{eq:experg})
this implies a rough proportionality between $M_\mathrm{ej}$ and
the nucleosynthesized $^{56}$Ni mass, but again with considerable
scatter associated with the relation between $E_\mathrm{exp}$  
and the iron-group nucleosynthesis, and the additional
(undetermined) factor $\epsilon\,\beta_\nu$ of Eq.~(\ref{eq:experg}).

\subsection{Kick dependence on the progenitor compactness}
\label{sec:progdept}

In this section we would like to address the question 
how the relevant ejecta mass, $M_\mathrm{ej}$, is 
connected to the properties of the core of the progenitor star.
$M_\mathrm{ej}$ as introduced in Sect.~\ref{sec:expdept}
is the expelled mass that carries the explosion asymmetries,
expressed by the asymmetry parameter $\alpha_\mathrm{ej}$,
during the time when the new-born NS is accelerated. We can
therefore write $M_\mathrm{ej}$ as the difference between
the stellar mass enclosed by the SN shock at the time when
the NS kick saturates and the initial (baryonic) mass of the 
compact remnant:
\begin{equation}
M_\mathrm{ej} = M_\mathrm{prog}(R_0) - M_\mathrm{NS,i}\,,
\label{eq:mej}
\end{equation}
where $R_0$ is the (average) shock radius at the considered
time. 

According to our present understanding of the neutrino-heating
mechanism, the onset of the explosion is favored when the
stalled SN shock enters into the collapsing, oxygen-enriched 
silicon shell, at which location the entropy per nucleon of the 
infalling matter exceeds a value of $s = 4\,k_\mathrm{B}$
per nucleon \citep{Ertletal2016,Sukhboldetal2016}. 
For this reason, a reasonable proxy of the initial NS mass is 
given by
\begin{equation}
M_\mathrm{NS,i} \approx M_\mathrm{prog}(s = 4) \,,
\label{eq:mns}
\end{equation}
i.e., by the progenitor mass that is enclosed by the 
radius where the entropy reaches $s = 4\,k_\mathrm{B}$ per 
nucleon.

It is more difficult to derive a useful estimate of 
$M_\mathrm{prog}(R_0)$. The NS kick velocity
approaches its final value when the acceleration
time scale becomes longer than the expansion time scale
of the anisotropic ejecta, in which case the gravitational
interaction between ejecta and NS become unimportant. This
requirement can be expressed by
\begin{equation}
t_\mathrm{acc} = \frac{v_\mathrm{NS}}{a_\mathrm{NS}} \gg
\frac{R_0}{\bar{v}_\mathrm{ej}} \,,
\label{eq:acccond}
\end{equation}
where 
\begin{equation}
\bar{v}_\mathrm{ej} = \sqrt{\frac{2E_\mathrm{kin}}{M_\mathrm{ej}}}
= \sqrt{\frac{2f_\mathrm{kin}E_\mathrm{exp}}{M_\mathrm{ej}}}
\label{eq:vej}
\end{equation}
is the average ejecta velocity and
\begin{equation}
a_\mathrm{NS} = \dot{v}_\mathrm{NS} \sim \alpha_\mathrm{ej}
\,\frac{G M_\mathrm{ej}}{R_0^2}
\label{eq:ans}
\end{equation}
is a crude measure of the NS acceleration. 
Using Eqs.~(\ref{eq:vns2a}), (\ref{eq:vej}), and (\ref{eq:ans})
in Eq.~(\ref{eq:acccond}) yields a condition for $R_0$:
\begin{equation}
R_0 \gg \frac{G}{2\,f_\mathrm{kin}}\,
\frac{M_\mathrm{ej}}{E_\mathrm{exp}}\,M_\mathrm{NS}\,.
\label{eq:r0inequality}
\end{equation}
Applying Eq.~(\ref{eq:experg}) we thus obtain
\begin{eqnarray}
R_0 &\gg& \frac{G}{2\,f_\mathrm{kin}\,\beta_\nu}\,
\epsilon^{-1}\,M_\mathrm{NS}
\label{eq:r0a} \\
&\sim& 200\,\mathrm{km}\,\,f_\mathrm{kin}^{-1}\,
\beta_\nu^{-1}\,\epsilon_5^{-1}\,
\left(\frac{M_\mathrm{NS}}{1.5\,M_\odot}\right)\,, 
\label{eq:r0b}
\end{eqnarray}
where in the second expression we used a representative
value of $\epsilon\sim 5$\,MeV/nucleon again and
once more 1.5\,$M_\odot$ as typical NS mass.
Since during the period of NS acceleration the kinetic energy is
usually only a minor fraction of the final explosion energy,
$f_\mathrm{kin}\sim 0.1$, and also $\beta_\nu$
drops with increasing shock radius to values 
around 0.1--0.2, Eq.~(\ref{eq:r0b})
suggests that typical values of $R_0$ are beyond 10,000\,km.
This is compatible with the numerical result that the NS 
acceleration in neutrino-driven explosions continues on a 
significant level for several seconds, during which period
the average ejecta velocity is a few 1000\,km\,s$^{-1}$
\citep{Schecketal2006,Wongwathanaratetal2010,Wongwathanaratetal2013}.

With the estimate of Eq.~(\ref{eq:r0b}) suggesting a rather generic
value of $R_0$, Eq.~(\ref{eq:mej}) implies that progenitors with
more massive cores enclosed by $R_0$ (i.e., progenitors with
bigger values of the core compactness $M_\mathrm{prog}(R_0)/R_0$)
tend to have larger ejecta masses $M_\mathrm{ej}$ and therefore 
higher explosion energies (see Eq.~\ref{eq:experg}).
For this reason such progenitors provide more favorable conditions
for higher NS kick velocities. We note in passing that progenitors
with these core properties also tend to produce more massive NSs.
A rough correlation of NS mass and explosion energy for 
neutrino-driven explosions was indeed reported by
\citet{Muelleretal2016b} (see figure~11 there).

The discussion outlined above is rather qualitative and
illuminates only basic dependences of the NS kick on the 
progenitor conditions. We will return to this topic in more
detail in Sect.~\ref{sec:numerical}.
In view of the linear relation of 
Eq.~(\ref{eq:experg}), however, the NS kick formula of 
Eq.~(\ref{eq:vns2b}) translates into Eq.~(\ref{eq:vns3}) without
requirement of any explicit knowledge of $M_\mathrm{ej}$. 

A similarly qualitative, though interesting, relation can
be derived by a different consideration. Nonspherical mass
distributions
in the postshock medium with anisotropic accretion downflows 
and buoyant outflows can continue to develop as long as the
postshock velocity, $v_\mathrm{pos}$,
is lower than the local escape velocity
\citep{MarekJanka2009,Mueller2015,Muelleretal2016b}. 
When the stellar
matter swept up by the outgoing NS shock expands faster than 
the escape speed, downflows to the NS will be quenched and
the spherically symmetric neutrino-driven wind will finally
push the asymmetric ejecta away from the NS, heralding the
phase when the NS acceleration ceases.
A very rough condition when the NS kick is determined
can therefore be coined by the relation
\begin{equation}
v_\mathrm{pos} \gtrsim v_\mathrm{esc}\,.
\label{eq:vposinequal}
\end{equation}
Here, we can approximately identify $v_\mathrm{pos}$ with the
average expansion velocity of the ejecta behind the shock
($v_\mathrm{pos}\sim \bar{v}_\mathrm{ej}$ with $\bar{v}_\mathrm{ej}$
from Eq.~\ref{eq:vej}), and the postshock escape velocity is given
by
\begin{equation}
v_\mathrm{esc} \sim \sqrt{\frac{2GM_\mathrm{prog}(R_0)}{R_0}}\,.
\label{eq:vesc}
\end{equation}
Introducing the dimensionless compactness parameter 
\citep{OConnorOtt2011}
\begin{equation}
\xi_0 = \frac{M_\mathrm{prog}(R_0)/M_\odot}{R_0/1000\,\mathrm{km}}\,,
\label{eq:xi0}
\end{equation}
we can write
\begin{equation}
v_\mathrm{esc} \sim \sqrt{2G\,\xi_0\,\frac{M_\odot}{10^8\,\mathrm{cm}}}\,.
\label{eq:vesc2}
\end{equation}
Eq.~(\ref{eq:vposinequal}) with this expression and 
with $v_\mathrm{pos}\sim\bar{v}_\mathrm{ej}$
from Eq.~(\ref{eq:vej}) leads to the condition
\begin{equation}
\frac{2f_\mathrm{kin}E_\mathrm{exp}}{M_\mathrm{ej}}
\gtrsim 2G\,\xi_0\,\frac{M_\odot}{10^8\,\mathrm{cm}}\,,
\end{equation}
which yields
\begin{equation}
\xi_0 \lesssim \frac{f_\mathrm{kin}}{G}\,
\frac{E_\mathrm{exp}}{M_\mathrm{ej}}\,\frac{10^8\,\mathrm{cm}}{M_\odot} 
= \frac{f_\mathrm{kin}}{G}\,\beta_\nu\,\epsilon\,
\frac{10^8\,\mathrm{cm}}{M_\odot}
\,,
\label{eq:xi0cond}
\end{equation}
where we have used Eq.~(\ref{eq:experg}) in the second 
expression on the rhs. Employing again our typical value for
$\epsilon$, we obtain
\begin{equation}
\xi_0 \lesssim 3.75\,f_\mathrm{kin}\,\beta_\nu\,\epsilon_5 \,.
\label{eq:xi0cond2}
\end{equation}
Since the compactness is a monotonically falling
quantity with increasing radius outside the ONeMg or
iron core (because the stellar 
density decline is steeper than $r^{-2}$ on average), 
Eq.~(\ref{eq:xi0cond2}) sets a lower limit to the 
distance from the center that must be reached by 
the shock for postshock asymmetries to freeze out and
for the NS kick to approach its terminal value.

The values of $f_\mathrm{kin}$, $\beta_\nu$, and 
$\epsilon$ vary only moderately between different 
progenitors during the NS acceleration phase of a
few seconds: $f_\mathrm{kin}$ is around 0.1 
(see figure~2 in \citealt{Ertletal2016} and figure~7 in
\citealt{Sukhboldetal2016}), and
$\beta_\nu$ is initially close to unity and later 
decreases continuously when not all stellar matter
swept up by the outgoing shock gets heated by neutrinos. 
Taking this for granted, Eq.~(\ref{eq:xi0cond2}) 
shows, again, a clear difference between progenitor
stars with a low core compactness and those with high values
of the core compactness: large NS kicks are disfavored 
by low compactness values, because a smaller ejecta mass
is neutrino-heated before the shock reaches radii
where the inequality condition of Eq.~(\ref{eq:xi0cond2})
is fulfilled. In particular
for low-mass stars, this condition is satisfied already
at the onset of the explosion. Such a low compactness
implies a fast acceleration of the explosion
and high ejecta velocities (of very little ejecta mass), 
leaving asymmetries in the
postshock layer little time to develop significant dipole 
amplitudes.

\subsection{Discussion of simulation results}
\label{sec:numerical}

Results of 2D and 3D simulations of neutrino-driven
explosions including the determination of NS kicks can
be found in the literature for constrained sets of
progenitor models. 
\citet{Schecketal2006} and \citet{Wongwathanaratetal2013} 
investigated stars in the birth-mass range of 15--20\,$M_\odot$,
\citet{Suwaetal2015} calculated explosions for ultra-stripped
SN~Ic progenitors \citep[see][]{Taurisetal2015}, and 
\citet{Gessner2014} \citep[also][]{GessnerJanka2017}
performed a systematic study for electron-capture SNe at
the low-mass end of core-collapse SN progenitors. 

Here we will set the results of these papers into the 
context of our discussion in Sects.~\ref{sec:expdept}
and \ref{sec:progdept}. Although the results are not finally
conclusive about possible variations of the relevant parameters
($\alpha_\mathrm{ej}$, $f_\mathrm{kin}$, $\beta_\nu$)
with progenitor and explosion conditions, they can still
provide first insights into some systematic
trends, which can be understood on grounds of the equations
derived above and can help to set constraints on the
remaining degrees of freedom. 

\citet{Schecketal2006} (figures~9, 11, C.1, and C.2 there) 
as well as \citet{Gessner2014} found very tight linear
relations between explosion energy $E_\mathrm{exp}$,
radial ejecta momentum $P_\mathrm{ej}$, and neutrino-heated 
ejecta mass $M_{\mathrm{ej},\nu}$, confirming the 
validity of Eq.~(\ref{eq:pej3}) (see also footnote~2 
of the present paper).
In fact, the relations for all investigated stellar models
with iron cores fall on top of each other, and the data 
for the ONeMg-core progenitor studied by \citet{Gessner2014}
connect continuously to the low-energy end of the Fe-core
results. This means that the
proportionality factors in the relations of 
Eq.~(\ref{eq:pej3}) are indeed independent of the 
considered progenitor (at least during the computed 
post-bounce period of one second) as we assumed in some 
arguments made in Sect.~\ref{sec:progdept}.

However, in both sets of simulations there is only a mild
positive correlation of the NS kick velocity with the explosion
energy \citep[see figure~9 in][]{Schecketal2006}, visible by
a slight shift of the ensemble distribution toward 
high-velocity cases for more energetic explosions. This is
confirmed by the 3D models of \citet{Wongwathanaratetal2013},
whose results also do not display clear trends with the 
explosion energy but rather indicate stronger
systematic differences between the considered progenitor 
models \citep[see figure~8 in][]{Wongwathanaratetal2013}.
One must caution, however, that the absence of an
unambiguous scaling with $E_\mathrm{exp}$ may be a 
modeling artifact connected to the use of a prescribed
neutrino luminosity from the NS core to trigger
the onset of the explosion. If this core luminosity is
overestimated relative to the accretion luminosity,
the detailed dynamics of the explosion and potentially 
also its asymmetry might be affected in an unrealistic way.
This may have been the case for the more energetic explosions
in \citet{Schecketal2006} and also for the cases with a
less rapidly contacting inner grid boundary to mimic
the time-dependence of the shrinking proto-NS radius, as
chosen by \citet{Wongwathanaratetal2013} to
minimize hydrodynamic time-step contraints in their 3D 
simulations. Fully self-consistent
calculations (not using a light-bulb contribution to
the neutrino emission) will be needed to clarify this aspect.

Moreover, the results of \citet{Schecketal2006} and
\citet{Wongwathanaratetal2013} show large case-to-case
variations of the NS kick velocity also for models with 
similar explosion energies. This velocity spread
can be understood by a large statistical scatter of
the asymmetry parameter $\alpha_\mathrm{ej}$ as visible
in figure~11 of \citet{Schecketal2006} and similarly in
the results of \citet{Gessner2014}. Such variations 
reflect the stochasticity of the growth of the explosion 
asymmetries that result from the chaotic interaction of 
several hydrodynamic instabilities (convective overturn and
Rayleigh-Taylor mass motions, SASI activity, Kelvin-Helmholtz
vortex motions at shear-flow interfaces) that play a role
in the postshock region on the way
to the onset of the explosion. The mean value and the
width of the $\alpha_\mathrm{ej}$ distribution exhibit
mild trends of decrease with higher explosion energies 
\citep[see][]{Schecketal2006,Wongwathanaratetal2013},
because more powerful explosions tend to develop
faster and thus to extenuate the merging of initially
higher-mode flow patterns to global asymmetries with 
dominant low-order spherical harmonics modes. This
mild systematic trend as well as the stochastic variations
of $\alpha_\mathrm{ej}$ mask the linear dependence
of $v_\mathrm{NS}$ on $E_\mathrm{exp}$ expressed by 
Eq.~(\ref{eq:vns3}). As mentioned above, the possible
influence of the core-neutrino light-bulb luminosity
should be kept in mind here. It is not clear how much
it might have affected the  mean and width of the 
$\alpha_\mathrm{ej}$ distibutions obtained by 
\citet{Schecketal2006} and \citet{Wongwathanaratetal2013}.
It is well conceivable that the statistical distribution
of $\alpha_\mathrm{ej}$ is fairly independent of the 
explosion energy and the mean value of the NS kick 
velocity in fully self-consistent simulations should
reflect the linear increase of $v_\mathrm{NS}$ 
with $E_\mathrm{exp}$ in
Eq.~(\ref{eq:vns3}) more prominently than visible in 
the present model sets.

While stochasticity plays an important role in all cases,
also the progenitor structure has a clear influence on the
NS kick as discussed in Sect.~\ref{sec:progdept}. Simulating
explosions of stellar models in the 15\,$M_\odot$ regime
for a wide range of explosion energies between less than
$0.2\times 10^{51}$\,erg and more than
$1.7\times 10^{51}$\,erg, \citet{Schecketal2006} found
values of $\alpha_\mathrm{ej}$ from basically zero to
about 0.33 with a mean of about 0.10--0.15. The 
average NS kicks were several 100\,km\,s$^{-1}$. In contrast,
low-mass progenitors with ONeMg cores and iron cores
are expected to explode with low energies of at most around
$10^{50}$\,erg 
\citep{Kitauraetal2006,Dessartetal2006,Jankaetal2012,Melsonetal2016}.
Varying the explosion energy for parametric, neutrino-driven
explosions of electron-capture SN models in 2D and 3D
between $\sim$$0.2\times 10^{50}$\,erg
and $\sim$$1.7\times 10^{50}$\,erg, \citet{Gessner2014} obtained
values of $\alpha_\mathrm{ej}$ ten times lower than 
\citet{Schecketal2006} (i.e., up to about 0.036) and NS kick
velocities of at most a few km\,s$^{-1}$ (the maximum kick
velocities were around 6\,km\,s$^{-1}$). Both the small explosion
energies (or small ejecta masses, see Eq.~\ref{eq:pej3})
and the small explosion asymmetries for the low-mass
progenitors are responsible
for a weak NS acceleration. The small explosion asymmetries
are caused by the rapid (quasi-spherical) 
development of the explosion and the fast expansion
of SN shock and ejecta. Both of these effects are favored
by the low compactness of the progenitors outside of the 
ONeMg or Fe-cores, which does not give hydrodynamic instabilities
and accretion downdrafts the chance to persist for a long 
period of time 
(see the discussion in connection to Eq.~\ref{eq:xi0cond2}).

\citet{Suwaetal2015} simulated explosions of bare CO stars to
mimic the progenitors of ultra-stripped SNe in binaries.
The compactness
values at an enclosed mass of 1.5\,$M_\odot$ for the
lower-mass cases of these ultra-stripped
models join the very small ones of the low-mass Fe-core and 
ONeMg-core progenitors, and for all cases they are 
considerably smaller than the compactness values of pre-collapse 
stars in the 15--20\,$M_\odot$ regime. On the basis of our
discussion we therefore expect NS kicks for ultra-stripped
SNe with iron cores to be intermediate between
those of electron-capture SNe and normal Fe-core SNe.
Indeed, \citet{Suwaetal2015} report NS kicks ranging from
$\sim$3\,km\,s$^{-1}$ to about 75\,km\,s$^{-1}$. This is 
compatible with the arguments presented here, but can only
be taken as suggestive, because \citet{Suwaetal2015} did
not explore large sets of models to account for stochastic 
variations of the explosion asymmetry.

\section{Possibilities for spin-kick alignment}
\label{sec:spinkick}

Observations of larger samples of radio pulsars and
of the compact remnants of Crab and Vela suggest an alignment 
of the kick direction and the spin (rotation) axis of 
many NSs
\citep[e.g.][]{Rankin2015,Noutsosetal2013,Kaplan2008,NgRomani2007,Laietal2001},
but see \citet{BrayEldridge2016}, who find no statistical 
evidence of a preferred kick orientation comparing NS velocity
distributions from observations and from their stellar 
population models for both single and binary star evolutionary 
pathways.

In order to connect this putative spin-kick alignment to the
internal SN dynamics, jets are often the first idea for
an explanation. However, this argument has several problems.
Jets are particularly inefficient in kicking the compact 
remnant (NS or BH), not only because they are collimated
outflows and as such usually contain only a small fraction of
the total ejecta mass and energy of the stellar explosion. In the 
case of relativistic jets the kick efficiency is additionally
reduced by the fact that the ejecta velocity is close to the
speed of light, $c$. In this case the momentum corresponding
to a jet energy $E_\mathrm{jet}$ is given by
$p_\mathrm{R} = E_\mathrm{jet}/c$. Because of the hugh number in
the denominator, this momentum of relativistic outflow is much
lower for a given energy than the momentum of Newtonian ejecta, 
where jet mass $M_\mathrm{jet}$ and energy $E_\mathrm{jet}$
are related by $p_\mathrm{N}=\sqrt{2M_\mathrm{jet}E_\mathrm{jet}}$.
For $M_\mathrm{jet} = 10^{-3}\,M_\odot$ and 
$E_\mathrm{jet} = 10^{51}\,$erg a formal comparison yields
$p_\mathrm{R} \approx 0.5\,p_\mathrm{N}$. 

The second reason why jets are an unlikely explanation for the
putative spin-kick alignment is the fact that in the far
majority of cases stellar cores at the time of collapse 
do not possess the huge amount of angular momentum (with
specific values in excess of $\sim$$10^{16}\,$cm$^2$\,s$^{-1}$)
needed for disk and jet formation \citep{Hegeretal2005}. 
Only in rare cases of
special stellar evolution scenarios, where the star avoids
massive mass loss and thus angular momentum loss
during the red-giant phase, the stellar
core at collapse may spin fast. This is an important
aspect in understanding the rarity of stellar death events
accompanied by long-duration gamma-ray bursts
\cite[e.g.][]{Levanetal2016}. It is also compatible with the
slow rotation rates observed for white dwarfs \citep{Kawaler2003}
and estimated for new-born NSs \citep{KaspiHelfand2002}, and
it is in accordance with
the efficient evolutionary spin down of stellar cores by a 
strong coupling between core and envelope as suggested by
asteroseismology \citep{Aerts2015}.

So the question remains, how the NS kick 
mechanism might lead to spin-kick aligment, if the 
observed alignment is a real effect and if
it points to an origin associated with the 
SN itself? Transfer of angular momentum to
the NS by stochastic accretion downflows during the 
post-bounce accretion phase in non-rotating progenitors
seems not to be able to achieve such an alignment but leads
to random orientations of NS spins relative to
the kick directions \citep{Wongwathanaratetal2013}.
The models in that paper, however, are subject to a 
number of constraints which prohibit too far reaching
conclusions. Besides the small sample of investigated
progenitors without rotation, the setup to compute 
artificially initiated neutrino-driven explosions
was not favorable for strong SASI activity, although in
self-consistent simulations phases of SASI sloshing and spiral
motions are found to be	prevalent for many models
\citep[e.g.][]{Hankeetal2013,Tamborraetal2014b,Kurodaetal2016}.

\begin{figure}
\includegraphics[width=\columnwidth]{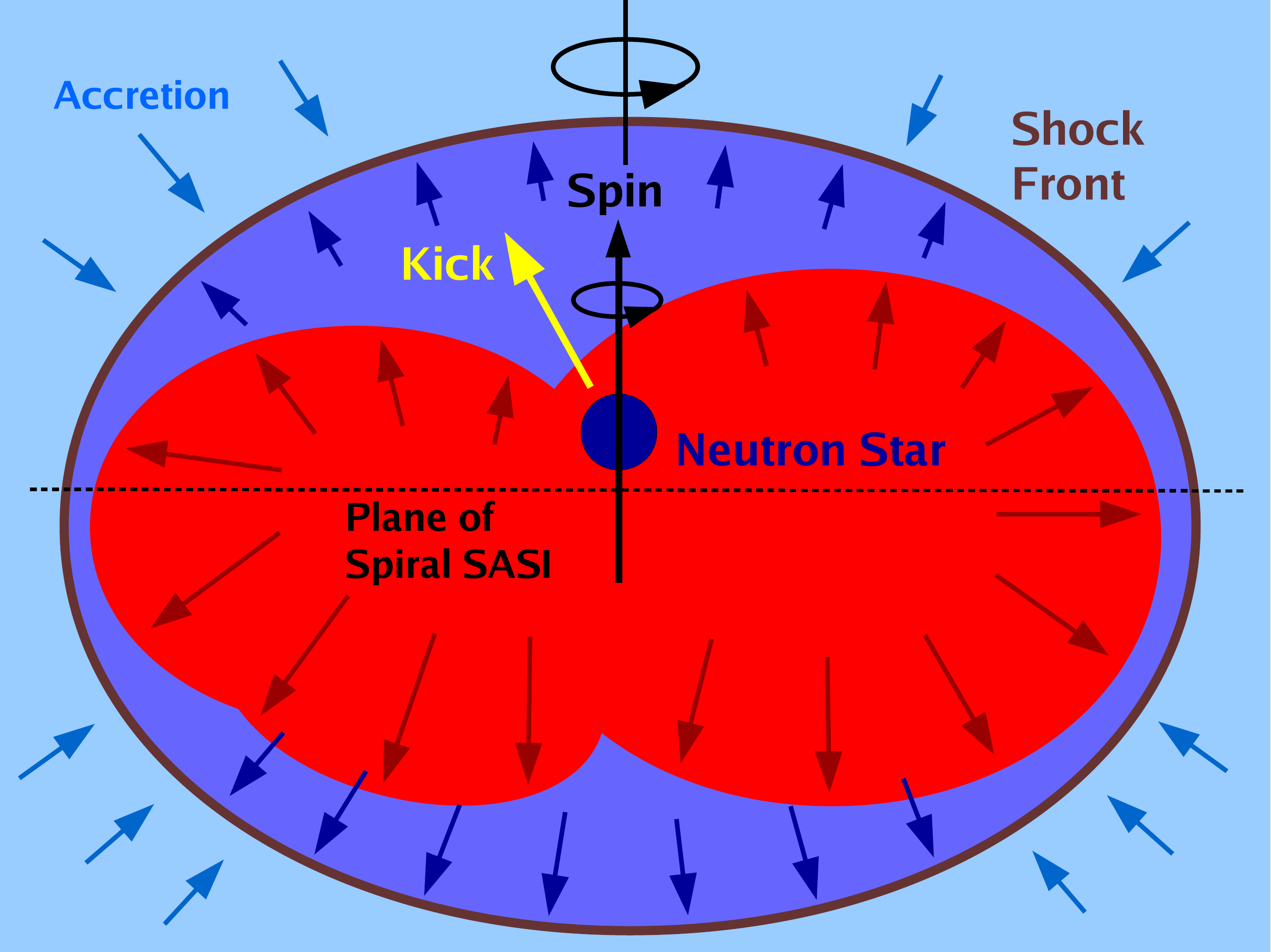}
%
\caption{
Spin-kick alignment resulting from a neutrino-driven explosion
launched from a phase of strong spiral-SASI activity. While the
explosion starts by equatorial expansion, the final NS kick is
determined by the slower mass ejection in the polar directions.
The red region symbolizes low-density, SASI deformed bubbles
of high-entropy, neutrino-heated matter, whereas the two inward
pointing ``noses'' in dark blue near the north pole and south
pole indicate the relics of long-lasting polar downflows of
shock-accreted low-angular-momentum matter. The NS is accelerated
by the gravitational attraction of the mass in these more
slowly expanding, dense regions. In the cartoon the NS is pulled
more strongly towards the northern direction and therefore
opposite to the (southern) hemisphere where the explosion is
more powerful.
The NS spin can be a consequence of the spiral SASI or can
be inherited from a rotating progenitor core, affected by
the angular momentum redistribution associated with the
spiral SASI.}
\label{fig:spinkick}
\end{figure}

Three scenarios that do not require to invoke extreme
assumptions for progenitor rotation
or NS magnetic fields, might account for NS kicks that
preferentially have small angles relative to the
NS spin axis.

First, angular momentum separation by spiral SASI motions 
in non-rotating or rotating progenitors has been
recognized to have important implications for NS rotation,
if the explosion is launched out of a SASI-active phase
\cite[e.g.][]{BlondinMezzacappa2007,Fernandez2010,GuiletFernandez2014,Kazeronietal2016,Kazeronietal2017}.
Such an explosion was, for example, observed in
a 3D model of a rotating 15\,$M_\odot$ progenitor by 
\citet{Jankaetal2016}. The blast wave in this case pushes
outward first around the equatorial plane, leaving 
long-lasting polar downflows, where accretion will be
quenched only at a later stage. In the case of a 
significant north-south asymmetry, these structures,
through hydrodynamic and gravitational forces in the
spirit of the gravitational tug-boat mechanism
\citep[see discussions in][]{Schecketal2006,Nordhausetal2010,Wongwathanaratetal2013},
must be expected to deflect the final NS kick towards
one of the poles even if the kick is close to the plane
of the spiral SASI motions at the onset of the explosion
(compare Fig.~\ref{fig:spinkick}). 

Second, large-scale velocity and density perturbations 
in the convective oxygen and silicon burning shells are
likely to have an effect on the asymmetry of the SN explosion 
\citep[e.g.][]{ArnettMeakin2011,CouchOtt2013,Couchetal2015,MuellerJanka2015,Chatzopoulosetal2016,Muelleretal2016,Mueller2016}. 
If stellar rotation facilitates a low-mode convective 
asymmetry between the north-polar and south-polar regions
of the burning shells, e.g.\ inflows on the one side and
outflows in the opposite hemisphere, the SN explosion might
develop a strong dipolar asymmetry along the rotation 
axis of the NS, similar to the explosion
asymmetry obtained by \citet{BurrowsHayes1996} as a result
of an artificially imposed lower-density wedge on one side
of the collapsing stellar iron core. A kick of the NS along its
spin axis and opposite to the direction of the stronger
explosion would be the consequence.

A third possibility for getting spin-kick alignment may
be connected to the dipolar neutrino-emission asymmetry
associated with the LESA (``self-sustained
lepton-number emission asymmetry'') phenomenon found
in 3D SN simulations recently \citep{Tamborraetal2014}.
If rotation is present in the stellar core, the LESA emission
dipole is oriented parallel to the rotation axis
\citep{Jankaetal2016}. The NS kick caused by the 
neutrino-emission dipole will therefore naturally be
aligned with the spin of the NS. Because of the small
dipole amplitude (of order per cent only) of the total
neutrino luminosity and the currently unknown duration
of the LESA emission dipole, present estimates suggest,
however, that the kick velocities thus obtained are
probably not higher than about 100\,km\,s$^{-1}$.

All of these potential scenarios for spin-kick alignment
might apply under different circumstances, but none of them
has been consolidated by self-consistent 
multi-dimensional SN simulations.
Long-time 3D calculations either of the post-bounce evolution 
(over periods of several seconds; scenario one and three) or
of the convective shell burning in rotating pre-SN stars on
their way to iron-core collapse (scenario two) for larger sets 
of progenitors are needed to assess the viability
of the described mechanisms for spin-kick alignment and the
magnitudes of the associated NS spins and kicks.

\section{Summary and conclusions}
\label{sec:conclusions}

In this paper we discussed how natal NS kicks by
the gravitational tug-boat mechanism in asymmetric SN
explosions depend on the properties of progenitor 
stars and explosions. Our approach was mostly on a 
didactic and conceptual level, referring to published
results in the literature, which, however, are not yet
conclusive in all aspects.
 
Our main result is Eq.~(\ref{eq:vns3}), which coins the
kick velocity as a function of the explosion energy, 
$E_\mathrm{exp}$, and of the momentum-asymmetry parameter
$\alpha_\mathrm{ej}$. By means of Eq.~(\ref{eq:emrelation}) 
the explosion energy can be replaced by the relevant ejecta
mass. The relevant ejecta mass is determined by explosion
models as the expelled mass behind the SN shock front
at the time when the NS kick asymptotes to its final value
and the (time-dependent) value of 
$\alpha_\mathrm{ej}$ is measured. It should not be confused
with the total ejecta mass of the SN, because the momentum
asymmetry for the total SN ejecta is usually not determined
by explosion models.

The main parameters, besides the NS mass, that govern the 
magnitude of the NS kick velocity are therefore the SN explosion 
energy and the momentum-asymmetry expressed by the parameter
$\alpha_\mathrm{ej}$. In the neutrino-driven mechanism,
according to our present understanding, $E_\mathrm{exp}$
might loosely correlate with the total SN ejecta
mass and the NS mass, but the scatter of individual cases
is considerable \citep{Muelleretal2016b}.

On grounds of our results we argued
that very small NS kick velocities can be expected for 
stars near the low-mass end of SN progenitors, which
possess very dilute envelopes around their degenerate
ONeMg or Fe-cores and are expected to explode with very
low energies of $\sim$$10^{50}$\,erg or less.
In addition, the rapid expansion of the SN shock prevents
the growth of hydrodynamic instabilities that lead to 
large dipolar asymmetry modes in the ejecta, for which reason 
$\alpha_\mathrm{ej}$ remains small, too.
The same conclusions can be drawn for ultra-stripped SNe with
(nearly) bare metal cores, which should leave behind NSs with
low or only moderate kick velocities, provided their core 
compactness is similarly low as that of the lowest-mass
unstripped core-collapse SN progenitors. 
For low-mass NSs that are born by progenitors
with small values of the core compactness near the
low-mass end of stars exploding as SNe, we therefore
do not only expect lower SN energies \citep[see also
figure~11 in][]{Muelleretal2016b} but also a tendency to
smaller kick velocities.

In contrast, higher average
natal NS kicks can be expected for explosions of more
massive SN progenitors, whose dense core environments 
enforce a longer 
delay of the onset of the explosion, thus permitting
the growth of low-mode hydrodynamic instabilities in the 
neutrino-heated postshock layer.
In such a situation, higher explosion asymmetries can
be obtained and also much larger amounts of mass are
involved in the neutrino-heating process, which favors 
higher explosion energies.
Both effects together lead to much stronger NS kicks.

Our conclusions are supported by larger sets of 
multi-dimensional hydrodynamic explosion simulations in several 
works
\citep{Schecketal2006,Wongwathanaratetal2013,Suwaetal2015,Gessner2014,GessnerJanka2017} and back up hypothetical low-kick scenarios discussed
in the literature involving electron-capture and ultra-stripped
SNe \citep{Podsiadlowski2004,Taurisetal2013,Taurisetal2015,Taurisetal2016}. 
We emphasize that ---in disagreement with arguments in the literature---
high NS kick velocities do not require a long shock stagnation
phase. The hydrodynamic instabilities in the postshock layer
develop within only $\sim$100--200\,ms
after core bounce. A corresponding delay of the shock runaway
is therefore sufficient for large-scale explosion asymmetries
to become possible. The models of \citet{Schecketal2006} and
\citet{Wongwathanaratetal2013} show most efficient NS acceleration
for cases where the explosion sets in fairly early after core bounce,
but the ejecta expand so slowly that the accretion-downflow
and mass-ejection asymmetries still grow afterwards. For such 
conditions the gravitational pull of these structures on the NS 
can continue on a high level for a long period of time to 
efficiently transfer momentum to the nascent NS. 

The simulations available in the literature, however, are not 
finally conclusive with respect to a possible dependence of the
momentum-asymmetry parameter $\alpha_\mathrm{ej}$ on the
explosion energy. The model results exhibit a tendency of 
showing the  highest NS kicks for moderate explosion energies
\citep[see figure~8 in][]{Wongwathanaratetal2013}, which
correlate with rather high neutrino-heated ejecta masses 
and large values of the ejecta asymmetry $\alpha_\mathrm{ej}$,
but not very rapid ejecta expansion. The underlying trend of
a reduced ejecta asymmetry with higher explosion energies might,
however, be a modeling artifact associated with the use of a 
light-bulb
prescription for the neutrino luminosity from the high-density
core of the NS, which dominates the accretion luminosity when
high-energy explosions are triggered in these parametric
simulations. Fully self-consistent SN models are needed to
determine the dependence of $\alpha_\mathrm{ej}$ on 
$E_\mathrm{exp}$. It is well possible that the statistical 
distribution of $\alpha_\mathrm{ej}$ is essentially independent
of the SN explosion energy for progenitors other than the 
discussed cases with lowest core compactness. 
More 3D hydrodynamic explosion models are also needed to 
determine the exact statistics (means and widths of the 
distribution functions) of $\alpha_\mathrm{ej}$ for
a wide range of progenitor stars.

Our main result, Eq.~(\ref{eq:vns3}) with the 
relation of Eq.~(\ref{eq:emrelation}), has some similarity
to a linear ansatz for the functional dependence of the NS
kick velocity on the ratio of the SN ejecta mass to the
NS mass recently proposed (but not physically explained)
by \citet{BrayEldridge2016}. These authors, however, used
the total mass of the SN ejecta, whose exact connection to
the NS kick is not well established on the theory side
as discussed here. Interestingly, an optimal fit of the 
two parameters of the linear function to the measured 
population-integrated NS kick distribution for a subset 
of pulsar observational data from \citet{Hobbsetal2005},
seems to require a constant floor value of the NS kick
velocity of more than 100\,km\,s$^{-1}$ even for vanishing
SN mass ejection. \citet{BrayEldridge2016} speculate that
this effect, if real, might be connected to a small 
neutrino-emission asymmetry such as it is, for example,
associated with the ``self-sustained
lepton-number emission asymmetry'' (LESA) that was recently 
discovered in 3D SN simulations by \citet{Tamborraetal2014}.
Estimates of the NS kick velocity associated with the
LESA asymmetry by \citet{Tamborraetal2014} are, however,
considerably lower than 100\,km\,s$^{-1}$. But a variety
of other neutrino-kick scenarios have been discussed in
the literature, mostly assuming non-standard neutrino
physics as well as strong magnetic fields in the NS
\citep[e.g.,][and references therein]{Bisnovatyi-Kogan1993,Kusenkoetal2008,SagertSchaffner2008}.
It has to be seen whether the interesting offset
of the NS kick function will survive further investigation
of observational data.

We point out that our discussion remains valid also if large-scale
perturbations in the convective burning shells of the pre-collapse
star play a role for the development and the asymmetry of 
neutrino-driven explosions \citep[e.g.,][]{ArnettMeakin2011,CouchOtt2013,Couchetal2015,MuellerJanka2015,Muelleretal2016,Mueller2016}. In this 
case the explosion asymmetry, expressed by our parameter
$\alpha_\mathrm{ej}$, may also depend on the large-scale
asphericities in the convective flows of the burning shells
\citep[see the results of][]{Mueller2016}, which could thus have
an important effect on the NS kick \citep{BurrowsHayes1996}. 
Presently, however, it is unclear for which progenitors and to
which extent such pre-collapse perturbations in the convective
Si- and/or O-shells have an impact on the SN explosion dynamics and
asymmetry, for which reason our discussion can just highlight a
potential relevance in principle.

\acknowledgments
The author is grateful to Thomas Tauris, Paulo Freire, Edward van den 
Heuvel, and Satoru Katsuda for 
inspiring discussions and to Thomas Tauris, Bernhard M\"uller, and 
Mathieu Renzo for very valuable comments on the manuscript.
This work was supported by the Deutsche Forschungsgemeinschaft through
the Excellence Cluster ``Universe'' EXC 153 and by the
European Research Council through grant ERC-AdG No.\ 341157-COCO2CASA.\\

\bibliographystyle{aasjournal}
\bibliography{kick}

\end{document}